\documentclass[11pt]{elsarticle}
\usepackage{amsmath,mathtools,amssymb}
\usepackage{graphicx}   
\usepackage{float} 
\usepackage{epstopdf}
\usepackage{subfigure}
\usepackage{color}
\usepackage{verbatim} 
\usepackage{booktabs}
\usepackage{bm}   
\usepackage{dutchcal}
\usepackage[left=3cm,right=3cm,top=2.5cm,bottom=2.5cm]{geometry}  
\usepackage{lineno,hyperref}
\usepackage{tablists} 
\usepackage{makecell} 
\usepackage{tikz}
\tikzstyle{arrow}     = [thick,->,>=stealth]
\usepackage{hyperref}
\usepackage{pstricks}

\usepackage{pstricks-add}
\hypersetup{
	colorlinks=ture,
	linkcolor=blue,
	filecolor=gray,
	urlcolor=blue,
	citecolor=blue}

\newtheorem{theorem}{Theorem}
\newtheorem{lemma}{Lemma}	
\newtheorem{proposition}{Proposition}

\begin{document}
	
	\begin{frontmatter}

	\title{3-form Yang-Mills based on 2-crossed modules}
	
	\author{Danhua Song\corref{cor1}}	
	\ead{danhua\_song@163.com}

	 \author{Kai Lou}
	 
	\author{Ke Wu}
	\ead{wuke@cnu.edu.cn}
	\author{Jie Yang}
	\ead{yangjie@cnu.edu.cn}
	\author{Fuhao Zhang}
	\cortext[cor1]{Corresponding author.}
	\address{School of Mathematical Sciences, Capital Normal University, Beijing 100048, China}

		\begin{abstract}
			In this paper, we study the higher Yang-Mills theory in the framework of higher gauge theory.
		    It was shown that the 2-form electromagnetism can be generalized to the 2-form Yang-Mills theory with the group $U(1)$ replaced by a crossed module of Lie groups.
		     To extend this theory to even higher structure, we develop a 3-form Yang-Mills theory with a 2-crossed module of Lie groups.
		     First, we give an explicit construction of non-degenerate symmetric $G$-invariant forms on the 2-crossed module of Lie algebras. Then, we derive the 3-Bianchi-Identities for 3-curvatures. Finally, we create a 3-form Yang-Mills action and obtain the corresponding field equations.
		\end{abstract}
	
		\begin{keyword}
			crossed module, 2-crossed module, 3-connection, 3-Bianchi-Identities, 3-form Yang-Mills
		\end{keyword}

	\end{frontmatter}

	  In modern mathematical physics a certain idea is very productive: the higher-dimensional extended objects are thought to be the basic constituents of matter and mediators of fundamental interactions. At present, higher gauge theory \cite{Baez.2010} seems to be the geometrically most promising technique to describe the dynamics of the higher-dimensional extended objects where gauge fields and field strengths are higher degree forms.
      Among them, the 2-gauge theory \cite{Baez2005HigherGT, Bar, JBUS} and the 3-gauge theory \cite{Faria_Martins_2011, Saemann:2013pca, doi:10.1063/1.4870640, Fiorenza:2012mr} have been studied deeply, whose higher algebraic structures and higher geometrical structures can be found in, e.g. \cite{TRMV, JBUS, U, ACJ, FH}.  In this article, we are concerned about the non-abelian higher gauge fields arising in a great deal of physical contexts, such as six dimensional superconformal field theory \cite{Saemann:2013pca}, quantum gravity \cite{TRMV}, string theory \cite{USJ} and M-theory \cite{ HSG, SP}, and so on.
	
	
	The electromagnetic theory is the simplest abelian gauge theory where the gauge field is described by the connection 1-form $A$ on a $U(1)$ bundle. 
	 When the charged particles are extended to $1$-dimensional charged strings, one can get the $2$-form electrodynamics \cite{HP} where the $2$-form gauge field can be described in terms of the $2$-form $B$ known as the Kalb-Ramond field \cite{Kalb:1974yc}.  The $2$-form gauge field $B$ is described by a connection on a $U(1)$ gerbe as the categorified version of a $U(1)$ bundle \cite{YZ}. 
	Furthermore, there exists a generalization known as $p$-form electrodynamics, where the $p$-form gauge fields play the role of gauge fields \cite{10.1007/BF01889624, EE, YTMM}. 
%
	When the group $U(1)$ is replaced by a nonabelian counterpart, one can obtain the Yang-Mills theory \cite{Mathai_2006, pittphilsci10008}. 
	Therefore, it raises a natural question of whether one can generalize Yang-Mills theory to a kind of higher-form Yang-Mills theory. 
	The main contribution of this work is to develop a general method for 2-form, 3-form Yang-Mills theories.

	In the related work \cite{2002hep.th....6130B}, Baez generalized the Yang-Mills theory to the 2-form Yang-Mills theory based on the $2$-form electrodynamics. The gauge group is replaced by a crossed module of Lie groups $\left(H,G;\alpha,\vartriangleright \right)$ \cite{ Beaz, Crans, Brown} satisfying the following two conditions:
		\begin{align}\label{gh}
			\alpha(g \vartriangleright h)&=g \alpha(h)g^{-1}, \ \ \ \ \ \ \ \forall g \in G, h \in H,
			\\
			\alpha(h_1)\vartriangleright h_2&=h_1 h_2 h^{-1}_1,  \ \ \ \ \ \ \ \forall h_1, h_2 \in H,
		\end{align} 
	where $\alpha: H  \longrightarrow G$ is a Lie group homomorphism and $\vartriangleright$ is a smooth action of the Lie group $G$ on the Lie group $H$ by automorphisms.
 A simple example of a Lie crossed module is $G = H = U(N) $ with $\alpha $ the identity map and the $ \vartriangleright $ the adjoint action.
	An associated differential crossed module $\left(\mathcal{h}, \mathcal{g}; \tilde{\alpha}, \widetilde{\vartriangleright} \right)$ can be constructed, where $\tilde{\alpha} : \mathcal{h} \longrightarrow \mathcal{g}$  is a Lie algebra map and $\widetilde{\vartriangleright}$ is a left action of the Lie algebra $\mathcal{g}$ of $G$ on the Lie algebra $\mathcal{h}$ of $H$ by derivations. Obviously, the differential version of the above example is $\mathcal{g} = \mathcal{h} = \mathcal{u} (N) $ with $\widetilde{\vartriangleright}$ being the adjoint action and $ \tilde{\alpha} $ being the identity map.
	We  represent the crossed module of Lie groups using squares of the form
	\begin{center}
		\begin{tikzpicture}[x=.5cm,y=.5cm]
		\draw[green,fill=green!10]  (0,2) -- (4,2) -- (4,-2) -- (0,-2) -- cycle;
		\node at (0, 0) {$g_1$};
		\node at (4, 0) {$g_2$};
		\node at (2, 0) {$h$};
		\end{tikzpicture}
	\end{center}
where $g_1$, $g_2 \in G$, $h \in H$ and $\alpha(h) = g_2 g^{-1}_1
$. Thus, we can calculate the crossed module with squares given by horizontally
\begin{center}
	\begin{tikzpicture}[x=.5cm,y=.5cm]
	\draw[green,fill=green!10]  (0,2) -- (4,2) -- (4,-2) -- (0,-2) -- cycle;
	\draw[green,fill=green!10]  (4,2) -- (8,2) -- (8,-2) -- (4,-2) -- cycle;
	\node at (0, 0) {$g_1$};
	\node at (4, 0) {$g_2$};
	\node at (2, 0) {$h_1$};
	\node at (8, 0) {$g_3$};
	\node at (6, 0) {$h_2$};
	\node at (10, 0) {=};
	
	\draw[green,fill=green!10]  (12,2) -- (16,2) -- (16,-2) -- (12,-2) -- cycle;
	
	\node at (12, 0) {$g_1$};
	\node at (16, 0) {$g_3$};
	\node at (14, 0) {$h_1 \circ h_2$};
	
	\node at (20,0) {$h_1 \circ h_2 := h_2 h_1$};
	\end{tikzpicture}
\end{center}
and vertically 
\begin{center}
	\begin{tikzpicture}[x=.5cm,y=.5cm]
	\draw[green,fill=green!10]  (0,4) -- (4,4) -- (4,0) -- (0,0) -- cycle;
	\draw[green,fill=green!10]  (0,0) -- (4,0) -- (4,-4) -- (0,-4) -- cycle;
	
	\node at (0, 2) {$g_1$};
	\node at (4, 2) {$g_2$};
	\node at (2, 2) {$h_1$};
	\node at (0, -2) {$g_3$};
	\node at (4, -2) {$g_4$};
	\node at (2, -2) {$h_2$};
	\node at (6, 0) {=};
	
	\draw[green,fill=green!10]  (8,2) -- (12,2) -- (12,-2) -- (8,-2) -- cycle;
	
	\node at (8, 0) {$g_1 g_3$};
	\node at (12, 0) {$g_2 g_4$};
	\node at (10, 0) {$h_1 \star h_2$};
	
	\node at (17.5, 0) {$h_1 \star h_2:= h_1 (g_1 \vartriangleright h_2)$.};
	\end{tikzpicture}
\end{center}
The two compositions are defined properly, since the squares resulting from the compositions satisfy 
\begin{align}
	\alpha(h_1 \circ h_2)&=g_3 g^{-1}_1, 
	\\
	\alpha(h_1 \star h_2)&=g_2 g_4 g^{-1}_3 g^{-1}_1,
\end{align}
by using the homomorphism $\alpha$ and the identity \eqref{gh}.
 Here, the horizontal composition of squares is denoted by a circle ``$\circ$'', and the vertical composition is denoted by a star ``$\star$''. 
 Squares admit horizontal and vertical inverses, defined by
\begin{center}
	\begin{tikzpicture}[x=.5cm,y=.5cm]
	\draw[green,fill=green!10]  (8,2) -- (12,2) -- (12,-2) -- (8,-2) -- cycle;
	\draw[green,fill=green!10]  (16,2) -- (20,2) -- (20,-2) -- (16,-2) -- cycle;
	
	\node at (8, 0) {$g_2$};
	\node at (12, 0) {$g_1$};
	\node at (10, 0) {$h^{-h}$};
	\node at (20, 0) {$g^{-1}_2$};
	\node at (16, 0) {$g^{-1}_1$};
	\node at (18, 0) {$h^{-v}$};
	\end{tikzpicture}
\end{center}
where $h^{-h} =h^{-1}$ and $h^{-v} =g_1^{-1} \vartriangleright h^{-1}$. The above definitions of inverses are reasonable since they satisfy 
	\begin{align}
	\alpha(h^{-h})=g_1 g^{-1}_2, \ \ \ \ \ \ \ \alpha(h^{-v})=g^{-1}_2 g_1.
	\end{align}
In addition, the vertical multiplication is the group multiplication of the group $G \ltimes H$, but the horizontal one is not.
There are some slightly different descriptions, see \cite{Baez.2010, HP, MORTON2020103548} for more details.
 Based on this algebraic structure, Baez constructed the 2-form Yang-Mills action and obtained the corresponding field equations. 
More information about the 2-form Yang-Mills theory can be found in \cite{2002hep.th....6130B}.
    In the special case when $H$ is trivial, the 2-form Yang-Mills equations reduce to the ordinary Yang-Mills equations with gauge group  $G$ \cite{pittphilsci10008}. When $G$ is trivial and $H = U(1)$, the 2-form Yang-Mills equations reduce to those of 2-form electromagnetism \cite{HP}.
	
	In another recent work \cite{Gastel}, Gastel gave two definitions of 2-Yang-Mills and 3-Yang-Mills theory which are linear if the good gauges are chosen. However, what we focus on here is different from his ideas. 
	In this paper, we extend the 2-form Yang-Mills theory of Baez's work to a 3-form Yang-Mills theory.
	Consequently, we need to introduce a new algebraic structure, known as 2-crossed module of Lie groups, which is described by three groups \cite{TRMV, Faria_Martins_2011, doi:10.1063/1.4870640, TRMV1, YHMNRY, TRMV3, Kamps20022groupoidEI, Mutlu1998FREENESSCF,Porter10thecrossed}. 
	We denote a 2-crossed module of Lie groups as $(L, H, G;\beta, \alpha, \vartriangleright, \bm{\left\{,\right\}})$ consisting of three Lie groups $L$, $H$ and $G$, and two group homomorphisms $\beta$ and $\alpha$
	$$L \stackrel{\beta}{\longrightarrow}H \stackrel{\alpha}{\longrightarrow}G,$$
		where $\alpha \beta =1$, and the smooth left action $\vartriangleright$ by automorphisms of $G$ on $L$ and $H$, and on itself by conjugation, i.e.
	\begin{equation}
	g \vartriangleright (e_1 e_2)=(g \vartriangleright e_1)(g \vartriangleright e_2),\ \ \ \ \ \ \ \  (g_1 g_2)\vartriangleright e = g_1 \vartriangleright (g_2 \vartriangleright e),
	\end{equation}
	for any $g, g_1, g_2\in G, e, e_1, e_2\in H $ or $L$, and a $G$-equivariant smooth function $ \bm{\left\{,\right\}} :H \times H \longrightarrow L $ called the Peiffer lifting, such that
	\begin{equation}
	g \vartriangleright \bm{\left\{} h_1, h_2 \bm{\right\}} = \bm{\left\{}g \vartriangleright h_1, g \vartriangleright h_2 \bm{\right\}},
	\end{equation}
	for any $g \in G$ and $h_1,h_2\in H$ and they satisfy certain relations. There is a left action of $H$ on $L$ by automorphisms $\vartriangleright' $ which is defined by
	\begin{equation}
	h \vartriangleright' l = l \bm{\left\{ } \beta (l)^{-1}, h \bm{\right\} }, \ \ \ \ \ \ \ \forall l \in L, h \in H.
	\end{equation}
	This together with the homomorphism $ \beta : L \longrightarrow H $ determines a crossed module $(L, H; \beta, \vartriangleright')$.
	 
	There is a differential 2-crossed module, also called a 2-crossed module of Lie algebras, corresponding to the 2-crossed module of Lie groups $(L, H, G;\beta, \alpha, \vartriangleright, \bm{\left\{,\right\}})$.
	This definition appeared in	\cite{E}. 
	We mark it with $(\mathcal{l},\mathcal{h}, \mathcal{g}; \tilde{\beta}, \tilde{\alpha}, \widetilde{\vartriangleright}, \left\{ , \right\})$ consisting of three Lie algebras $\mathcal{l}$, $\mathcal{h}$ and $\mathcal{g}$ respectively corresponding to the three Lie groups $L$, $H$ and $G$, and two Lie algebra maps $\tilde{\beta}$ and $\tilde{\alpha}$
    	$$\mathcal{l}\stackrel{\tilde{\beta}}{\longrightarrow} \mathcal{h} \stackrel{\tilde{\alpha}}{\longrightarrow} \mathcal{g},$$
	where $ \tilde{\alpha}\tilde{\beta} = 0$, and left action $\widetilde{\vartriangleright} $ of $\mathcal{g}$ on $\mathcal{l}, \mathcal{h}$ and $\mathcal{g}$ by automorphisms, and a $\mathcal{g}$-equivariant bilinear map $\left\{ , \right\}:\mathcal{h} \times \mathcal{h} \longrightarrow \mathcal{l}$ called the Peiffer lifting, such that
	\begin{equation}\label{12}
	X \widetilde{\vartriangleright} \left\{ Y_1,Y_2\right\} = \left\{X \widetilde{\vartriangleright} Y_1,Y_2\right\} + \left\{Y_1, X \widetilde{\vartriangleright} Y_2\right\}, \ \ \ \ \ \ \ \forall X \in \mathcal{g}, Y_1,Y_2 \in \mathcal{h},
	\end{equation}
	and they satisfy appropriate conditions \cite{Faria_Martins_2011}. Analogously to the 2-crossed module of Lie groups case, there is a left action $\widetilde{\vartriangleright}'$ of $\mathcal{h}$ on $\mathcal{l}$ which is defined by
	\begin{equation}\label{YZ}
	Y\widetilde{\vartriangleright}'Z= -\left\{ \tilde{\beta}(Z),Y\right\}, \ \ \ \ \ \ \ \forall Y\in \mathcal{h}, Z \in \mathcal{l}.
	\end{equation}
	This together with the homomorphism $\tilde{\beta} : \mathcal{l} \longrightarrow \mathcal{h}$ defines a differential crossed module $(\mathcal{l}, \mathcal{h}; \tilde{\beta}, \widetilde{\vartriangleright}')$.

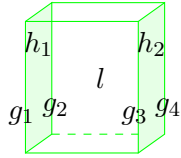
\begin{figure}[H]
		\begin{center}
			\begin{tikzpicture}[x=.5cm,y=.5cm]
				\draw[green,fill=green!10] (14.5,1.5) -- (15.2,2) -- (15.2,-1.5) -- (14.5,-2) -- cycle;
				\draw[green,fill=green!10] (17.5,1.5) -- (18.2,2) -- (18.2,-1.5) -- (17.5,-2) -- cycle;
				
				\draw[green,line cap=miter] (14.5,1.5) -- (17.5,1.5);
				\draw[green,dashed] (15.2,-1.5) -- (17.5,-1.5);
				\draw[green,line cap=miter] (15.2,2) -- (18.2,2);
				\draw[green,line cap=miter] (14.5,-2) -- (17.5,-2);
				
				\node at (16.5, 0) {$l$};
				\node at (14.4, -1) {$g_1$};
				\node at (15.3, -0.8) {$g_2$};
				\node at (17.4, -1) {$g_3$};
				\node at (18.3, -0.8) {$g_4$};
				\node at (14.85, 0.9) {$h_1$};
				\node at (17.85, 0.9) {$h_2$};
			\end{tikzpicture}
		\end{center}
		\caption{Cube of 2-crossed module}\label{cube}
	\end{figure}

	In consideration of the crossed module of Lie groups $(L, H; \beta, \vartriangleright')$ satisfying
		\begin{align}
		&	\beta (h \vartriangleright' l) = h \beta (l) h^{-1},  \ \ \ \ \ \ \ \forall h \in H, l \in L,
			\\
		&	\beta (l) \vartriangleright' l' = l l' l^{-1}, \ \ \ \ \ \ \ \ \ \ \ \ \forall l , l' \in L,
	\end{align}
and its calculations with squares as shown above, we represent  the 2-crossed module of groups by using cubes as Figure \ref{cube}, where $g_1$, $g_2$, $g_3$, $g_4 \in G$, $h_1$, $h_2 \in H$, $l \in L$ and $\beta (l) = h_2 h^{-1}_1$.

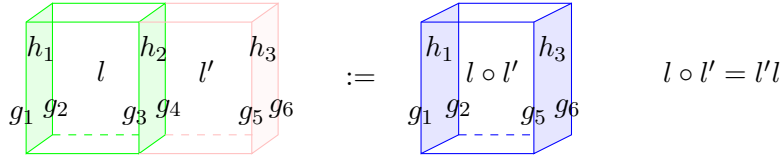
\begin{figure}[H]
		\begin{center}
			\begin{tikzpicture}[x=.5cm,y=.5cm]
				\draw[green,fill=green!10] (14.5,1.5) -- (15.2,2) -- (15.2,-1.5) -- (14.5,-2) -- cycle;
				\draw[green,fill=green!10] (17.5,1.5) -- (18.2,2) -- (18.2,-1.5) -- (17.5,-2) -- cycle;
				\draw[pink,fill=pink!10] (20.5,1.5) -- (21.2,2) -- (21.2,-1.5) -- (20.5,-2) -- cycle;
				
				\draw[green,line cap=miter](14.5,1.5) -- (17.5,1.5);
				\draw[green,line cap=miter](15.2,2) -- (18.2,2);
				\draw[green,dashed](15.2,-1.5) -- (17.5,-1.5);
				\draw[green,line cap=miter](14.5,-2)  -- (17.5,-2) ;
				\draw[pink,line cap=miter](17.5,1.5)  -- (20.5,1.5) ;
				\draw[pink,line cap=miter](18.2,2)  -- (21.2,2) ;
				\draw[pink,dashed](18.2,-1.5)  -- (20.6,-1.5) ;
				\draw[pink,line cap=miter](17.5,-2)  -- (20.5,-2) ;
				
				\node  at (16.5, 0.2) {$l$};
				\node  at (14.4, -1) {$g_1$};
				\node  at (15.3, -0.8) {$g_2$};
				\node at (17.4, -1) {$g_3$};
				\node at (18.3, -0.8) {$g_4$};
				\node  at (14.9, 0.8) {$h_1$};
				\node at (17.9, 0.8) {$h_2$};
				\node at (20.5, -1) {$g_5$};
				\node at (21.3, -0.8) {$g_6$};
				\node at (20.8, 0.8) {$h_3$};
				\node at (19.3, 0.2) {$l'$};
				
				\node  at (23.4, 0) {:=};
				
				\draw[blue,fill=blue!10] (25,1.5) -- (26,2) -- (26,-1.5) -- (25,-2) -- cycle;	
				\draw[blue,fill=blue!10] (28,1.5) -- (29,2) -- (29,-1.5) -- (28,-2) -- cycle;	
				
				\draw[blue,line cap=miter](25,1.5) -- (28,1.5);
				\draw[blue,line cap=miter](26,2) -- (29,2);
				\draw[blue,dashed](26,-1.5)  -- (28,-1.5);
				\draw[blue,line cap=miter](25,-2)  -- (28,-2) ;
				
				\node at (25, -1) {$g_1$};
				\node at (26, -0.8) {$g_2$};
				\node at (28, -1) {$g_5$};
				\node at (28.9, -0.8) {$g_6$};
				\node at (25.5, 0.8) {$h_1$};
				\node at (28.5, 0.8) {$h_3$};
				\node at (26.9, 0.2) {$l \circ l'$};
				\node at (33.0, 0.2) { $l \circ l' = l' l$ };
			\end{tikzpicture}
		\end{center}
		\caption{The horizontal composition}\label{figure 2}
	\end{figure}

There are two types of composition between these cubes. The first one is the  horizontal composition given by Figure \ref{figure 2}, which satisfies
	\begin{align}
\beta(l \circ l')=\beta(l') \beta(l)=h_3 h^{-1}_1
	\end{align}
	 by using $\beta(l)=h_2 h^{-1}_1$ and $\beta(l')=h_3 h^{-1}_2$.  And the second one is the vertical composition given by Figure \ref{figure 3}, which satisfies 
	 \begin{align}
	 \beta(l \star l')=\beta(l(h_1 \vartriangleright' l'))=h_2  h_4 h^{-1}_3 h^{-1}_1
	 \end{align}
	 by using $\beta(l)=h_2 h^{-1}_1$, $\beta(l')=h_4 h^{-1}_3$ and $\beta(h_1 \vartriangleright' l')=h_1 \beta(l')h^{-1}_1$. 
		Correspondingly, there are two different inverses under the above two compositions.
		The horizontal inverse is defined by Figure  \ref{figure 5} satisfying
		\begin{align}
	\beta(l^{-h})=h_1 h^{-1}_2,
		\end{align}
		 and the vertical one is given by Figure \ref{figure 6} satisfying
		 	\begin{align}
		 	\beta(l^{-v})=h^{-1}_2 h_1.
		 \end{align}
	Here, we restrict ourselves to the particular 2-crossed module of groups in which $(H,G; \alpha, \vartriangleright)$ is also a crossed module.
	\begin{figure}[H]
		\begin{center}
			\begin{tikzpicture}[x=.5cm,y=.5cm]
				\draw[green,fill=green!10] (14.5,1.5) -- (15.2,2) -- (15.2,-1.5) -- (14.5,-2) -- cycle;
				\draw[green,fill=green!10] (17.5,1.5) -- (18.2,2) -- (18.2,-1.5) -- (17.5,-2) -- cycle;
				\draw[pink,fill=pink!10] (14.5,-2) -- (15.2,-1.5) -- (15.2,-5) -- (14.5,-5.5) -- cycle;
				\draw[pink,fill=pink!10] (17.5,-2) -- (18.2,-1.5) -- (18.2,-5) -- (17.5,-5.5) -- cycle;
				
				\draw[green,line cap=miter](14.5,1.5) -- (17.5,1.5);
				\draw[green,line cap=miter](15.2,2) -- (18.2,2);
				\draw[green,dashed](15.2,-1.5) -- (17.5,-1.5);
				\draw[green,line cap=miter](14.5,-2)  -- (17.5,-2) ;
				\draw[pink,dashed](15.2,-5)  -- (17.5,-5) ;
				\draw[pink,line cap=miter](14.5,-5.5)  -- (17.5,-5.5) ;
				
				\node at (16.5, 0) {$l$};
				\node at (15.0, 1.9) {$g_1$};
				\node at (15.0, -1.7) {$g_2$};
				\node at (15.0, -5.3) {$g_5$};
				\node at (18.0, 1.9) {$g_3$};
				\node at (18.0, -1.7) {$g_4$};
				\node at (18.0, -5.3) {$g_6$};
				\node at (14.9, 0.2) {$h_1$};
				\node at (17.9, 0.2) {$h_2$};
				
				\node  at (16.5, -3.3) {$l'$};
				\node at (14.9, -3.7) {$h_3$};
				\node at (17.9, -3.7) {$h_4$};
				\node  at (20.5, -1.5) {:=};
				
				\draw[blue,fill=blue!10] (22.5,0) -- (23.5,0.5) -- (23.5,-3) -- (22.5,-3.5) -- cycle;
				\draw[blue,fill=blue!10] (25.5,0) -- (26.5,0.5) -- (26.5,-3) -- (25.5,-3.5) -- cycle;
				
				\draw[blue,line cap=miter](22.5,0)-- (25.5,0);
				\draw[blue,line cap=miter](23.5,0.5) -- (26.5,0.5);
				\draw[blue,dashed](23.5,-3) -- (25.5,-3);
				\draw[blue,line cap=miter](22.5,-3.5)  -- (25.5,-3.5) ;
				
				\node at (24.7, -1.0) {$l \star l'$};
				\node at (23, 0.4) {$g_1 $};
				\node at (23 , -3.1) {$g_5$};
				\node at (26, 0.4) {$g_3 $};
				\node at (26, -3.1) {$g_6$};
				\node at (23, -1.5) {$h_3 \circ h_1 $};
				\node at (26, -1.5) {$h_4 \circ h_2$};
				\node at (32, 0.1) {$h_3 \circ h_1 = h_1 h_3$};
				\node at (32, -1.5) {$h_4 \circ h_2 = h_2 h_4$};
				\node at (32, -3.1) {$l \star l' = l(h_1 \vartriangleright' l')$};
			\end{tikzpicture} 
		\end{center}
	\caption{The vertical composition} \label{figure 3}
	\end{figure}

	\begin{figure}[H]
		\begin{center}
			\begin{minipage}[t]{0.4\textwidth}
				\begin{tikzpicture}[x=.5cm,y=.5cm]
					\draw[green,fill=green!10] (21.5,1.5) -- (22.2,2) -- (22.2,-1.5) -- (21.5,-2) -- cycle;
					\draw[green,fill=green!10] (24.5,1.5) -- (25.2,2) -- (25.2,-1.5) -- (24.5,-2) -- cycle;
					
					\draw[green,line cap=miter] (21.5,1.5) -- (24.5,1.5);
					\draw[green,dashed] (22.2,-1.5) -- (24.5,-1.5);
					\draw[green,line cap=miter] (22.2,2) -- (25.2,2);
					\draw[green,line cap=miter] (21.5,-2) -- (24.5,-2);
					
					\node at (23.5, 0) {$l^{-h}$};
					\node at (21.4, -1) {$g_3$};
					\node at (22.3, -0.8) {$g_4$};
					\node at (24.4, -1) {$g_1$};
					\node at (25.3, -0.8) {$g_2$};
					\node at (24.85, 0.9) {$h_1$};
					\node at (21.85, 0.9) {$h_2$};
					\node at (29.5, 0) {$l^{-h}=l^{-1}$};
				\end{tikzpicture}
				\caption{The horizontal inverse}
				\label{figure 5}
			\end{minipage}
		\end{center}
	\end{figure}
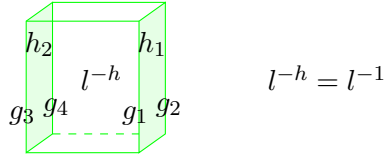
	
	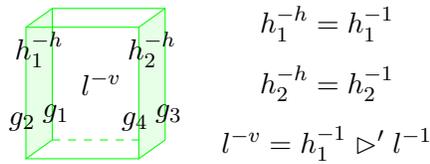
\begin{figure}[H]
		\begin{center}
			\begin{minipage}[t]{0.4\textwidth}
				\begin{tikzpicture}[x=.5cm,y=.5cm]
					\draw[green,fill=green!10] (15.5,1.5) -- (16.2,2) -- (16.2,-1.5) -- (15.5,-2) -- cycle;
					\draw[green,fill=green!10] (18.5,1.5) -- (19.2,2) -- (19.2,-1.5) -- (18.5,-2) -- cycle;
					
					\draw[green,line cap=miter] (15.5,1.5) -- (18.5,1.5);
					\draw[green,dashed] (16.2,-1.5) -- (18.5,-1.5);
					\draw[green,line cap=miter] (16.2,2) -- (19.2,2);
					\draw[green,line cap=miter] (15.5,-2) -- (18.5,-2);
					
					\node at (17.5, 0) {$l^{-v}$};
					\node at (15.4, -1) {$g_2$};
					\node at (16.3, -0.8) {$g_1$};
					\node at (18.4, -1) {$g_4$};
					\node at (19.3, -0.8) {$g_3$};
					\node at (15.85, 0.9) {$h^{-h}_1$};
					\node at (18.85, 0.9) {$h^{-h}_2$};
					\node at (23.5, 1.6) {$h^{-h}_1 = h^{-1}_1$};
					\node at (23.5, 0) {$h^{-h}_2 = h^{-1}_2$};
					\node at (23.5, -1.6) {$l^{-v}= h^{-1}_1 \vartriangleright' l^{-1}$};
				\end{tikzpicture}
				\caption{The vertical inverse}
				\label{figure 6}
			\end{minipage}
		\end{center}
	\end{figure}
	
%
%

	In order to construct the 3-form Yang-Mills action, we introduce some kind of $G$-invariant bilinear forms on the 2-crossed module of Lie algebras. More precisely speaking, we generalize the $G$-invariant bilinear forms on the  crossed module of Lie algebras \cite{Martins:2010ry, Radenkovi__2019}. 
    Firstly, we generalize the  mixed relations of $(H, G;\alpha, \vartriangleright)$  and $(\mathcal{h}, \mathcal{g};\tilde{\alpha}, \widetilde{\vartriangleright})$ to the mixed relations of
	  $(L, H, G;\beta, \alpha, \vartriangleright, \bm{\left\{,\right\}})$  and $(\mathcal{l},\mathcal{h}, \mathcal{g}; \tilde{\beta}, \tilde{\alpha}, \widetilde{\vartriangleright}, \left\{ , \right\})$.
    When $(H, G; \alpha, \vartriangleright)$ is a crossed module of Lie groups, there are two mixed relations \cite{Martins:2010ry}
   induced by the action via $\vartriangleright$ on $H$ and also denoted by $\vartriangleright$
   \begin{align}
   \tilde{\alpha} (g \vartriangleright Y) = g \tilde{\alpha}(Y) g^{-1}, \ \ \ \ \ \ \ \forall g \in G, Y \in \mathcal{h},
   \end{align}
   \begin{align}\label{mixed1}
   \alpha (h) \vartriangleright Y = h Y h^{-1}, \ \ \ \ \ \ \ \forall h \in H, Y \in \mathcal{h}.
   \end{align}
   Besides, $(L, H; \beta, \vartriangleright')$ is also a crossed module of Lie groups.
   Thus there are also the mixed relations, corresponding to the crossed module of Lie algebras $(\mathcal{l}, \mathcal{h}; \tilde{\beta}, \widetilde{\vartriangleright}')$, induced by the action via $\vartriangleright'$ on $L$ and also denoted by $\vartriangleright'$
   \begin{align}
   \tilde{\beta} (h \vartriangleright' Z) = h \tilde{\beta} (Z) h^{-1},  \ \ \ \ \ \ \ \forall h \in H, Z \in \mathcal{l},
   \end{align}
   \begin{align}\label{mixed2}
   \beta (l) \vartriangleright' Z = l Z l^{-1}, \ \ \ \ \ \ \ \forall l \in L, Z \in \mathcal{l}.
   \end{align}
	
	A symmetric non-degenerate $G$-invariant form in $(\mathcal{l},\mathcal{h}, \mathcal{g}; \tilde{\beta}, \tilde{\alpha}, \widetilde{\vartriangleright}, \left\{ , \right\})$ is given by a triple of non-degenerate symmetric bilinear forms $\langle,\rangle_\mathcal{g}$ in $\mathcal{g}$ , $\langle,\rangle_\mathcal{h}$ in $\mathcal{h}$ and $\langle,\rangle_\mathcal{l}$ in $\mathcal{l}$ such that
	\begin{enumerate}
		\item  $\langle,\rangle_\mathcal{g}$ is $G$-invariant, i.e.
		$$ \langle gXg^{-1}, gX'g^{-1} \rangle_\mathcal{g}= \langle X,X'\rangle_\mathcal{g}, \ \ \ \forall g \in G, X, X' \in \mathcal{g};$$
		\item  $\langle,\rangle_\mathcal{h}$ is $G$-invariant, i.e.
		$$ \langle g\vartriangleright Y,  g\vartriangleright Y' \rangle_\mathcal{h}= \langle Y,Y'\rangle_\mathcal{h}, \ \ \ \ \ \ \ \forall g\in G,Y,Y'\in \mathcal{h};$$
		\item  $\langle,\rangle_\mathcal{l}$ is $G$-invariant, i.e.
		$$ \langle g\vartriangleright Z,  g\vartriangleright Z' \rangle_\mathcal{l}= \langle Z,Z'\rangle_\mathcal{l}, \ \ \ \ \ \ \ \forall g\in G,Z,Z'\in \mathcal{l}.$$
	\end{enumerate}
Note that $\langle,\rangle_\mathcal{h}$ is necessarily $H$-invariant. Since
$$\langle h Y h^{-1}, h Y' h^{-1} \rangle_{\mathcal{h}} = \langle \alpha(h) \vartriangleright Y, \alpha(h) \vartriangleright Y' \rangle_{\mathcal{h}} = \langle Y, Y'\rangle_{\mathcal{h}}, \ \ \ \ \forall h \in H, Y, Y' \in \mathcal{h},$$
where we have used the mixed relation \eqref{mixed1}.
In the case when the Peiffer lifting or the map $\beta$ is trivial consequently, $\langle,\rangle_\mathcal{l}$ is $H$-invariant, i.e.
$$ \langle h\vartriangleright' Z, h \vartriangleright' Z' \rangle_\mathcal{l} = \langle Z,Z'\rangle_\mathcal{l}, \ \ \ \ \forall h\in H, Z,Z' \in \mathcal{l}.$$
Besides, $\langle,\rangle_\mathcal{l}$ is necessarily $L$-invariant based on the $H$-invariance of $\langle,\rangle_\mathcal{l}$, i.e.
$$\langle lZl^{-1}, lZ'l^{-1} \rangle_\mathcal{l} = \langle \beta (l) \vartriangleright' Z, \beta (l) \vartriangleright' Z' \rangle_\mathcal{l} = \langle Z,Z'\rangle_\mathcal{l}, \ \ \ \ \forall l\in L, Z,Z' \in \mathcal{l},$$
where the mixed relation \eqref{mixed2} is used. See \cite{Radenkovi__2019} for details.
	
	There are no compatibility conditions among the bilinear forms $\langle,\rangle_\mathcal{l}$,  $\langle,\rangle_\mathcal{h}$ and  $\langle,\rangle_\mathcal{g}$. 
	From the well-known fact that any representation of $G$ can be made unitary if $G$ is a compact group, one can get the following Lemma.
	\begin{lemma}
		Let  $(L, H, G;\beta, \alpha, \vartriangleright, \bm{\left\{,\right\}})$  be a 2-crossed module of Lie groups with the group $G$ being compact in the real case, or having a compact real form in the complex case. Then one can construct $G$-invariant symmetric non-degenerate bilinear forms $\langle,\rangle_\mathcal{g}$, $\langle,\rangle_\mathcal{h}$ and $\langle,\rangle_\mathcal{l}$ in the associated differential 2-crossed module  $(\mathcal{l},\mathcal{h}, \mathcal{g}; \tilde{\beta}, \tilde{\alpha}, \widetilde{\vartriangleright}, \left\{ , \right\})$. Furthermore these forms can be chosen to be positive definite.
	\end{lemma}

These invariance conditions imply that:
\begin{align}\label{XXX}
\langle [X,X'], X''\rangle_\mathcal{g}=-\langle X',[X,X'']\rangle_\mathcal{g},
\end{align}
\begin{align}\label{YYY}
\langle [Y,Y'], Y''\rangle_\mathcal{h}=-\langle Y',[Y,Y'']\rangle_\mathcal{h},
\end{align}
\begin{align}\label{ZZZ}
\langle [Z,Z'],Z''\rangle_\mathcal{l}=-\langle Z',[Z,Z'']\rangle_\mathcal{l}.
\end{align}

One can define two bilinear antisymmetric maps $\sigma:\mathcal{h} \times \mathcal{h} \longrightarrow \mathcal{g}$ by the rule:
\begin{align}\label{Qw1}
\langle \sigma(Y,Y'), X\rangle_\mathcal{g}=-\langle Y, X\widetilde{\vartriangleright} Y'\rangle_\mathcal{h}, \ \ \ \ \ \ \ \forall X \in \mathcal{g}, Y,Y' \in \mathcal{h},
\end{align}
and   $\kappa:\mathcal{l} \times \mathcal{l} \longrightarrow \mathcal{g}$ by the rule:
\begin{align}\label{Qw2}
\langle \kappa(Z,Z'), X\rangle_\mathcal{g}=-\langle Z, X\widetilde{\vartriangleright} Z'\rangle_\mathcal{l},  \ \ \ \ \ \ \ \forall X \in \mathcal{g}, Z,Z' \in \mathcal{l}.
\end{align}
Especially, when $\mathcal{h}=\mathcal{l}=\mathcal{g}$ and the bilinear form is the Killing form i.e. $\langle -, - \rangle_{\mathcal{g}} = K( -, - )$, the above definitions will become very natural as follows:
$$\sigma(Y, Y')=[Y, Y'],\ \ X\widetilde{\vartriangleright} Y'= [X, Y'], \ \ \kappa(Z, Z')=[Z, Z'],\ \ X\widetilde{\vartriangleright} Z'= [X, Z'],$$
and \eqref{Qw1} and \eqref{Qw2}will become Killing identities
$$K([Y, Y'], X) + K(Y, [X, Y'])=0,$$$$K([Z, Z'], X) + K(Z, [X, Z'])=0.$$ 
Due to $\sigma(Y',Y)=-\sigma(Y,Y')$ and $\kappa(Z',Z)=-\kappa(Z,Z')$, then one has
\begin{align}\label{YXY'}
\langle Y, X\widetilde{\vartriangleright} Y'\rangle_\mathcal{h}=-\langle Y', X\widetilde{\vartriangleright} Y \rangle_\mathcal{h}=-\langle X\widetilde{\vartriangleright} Y, Y' \rangle_\mathcal{h},
\end{align}
\begin{align}\label{ZXZ'}
\langle Z, X\widetilde{\vartriangleright} Z'\rangle_\mathcal{l}=-\langle Z', X\widetilde{\vartriangleright} Z \rangle_\mathcal{l}=-\langle X\widetilde{\vartriangleright} Z, Z' \rangle_\mathcal{l}.
\end{align}

Further, one needs to define two bilinear maps $\eta_1: \mathcal{l} \times \mathcal{h} \longrightarrow \mathcal{h}$ and $\eta_2: \mathcal{l} \times \mathcal{h} \longrightarrow \mathcal{h}$ by the rule:
\begin{align}\label{YY'Z}
\langle \left\{ Y, Y' \right\}, Z \rangle_\mathcal{l} = -\langle Y', \eta_1(Z, Y) \rangle_\mathcal{h}= -\langle Y, \eta_2(Z, Y') \rangle_\mathcal{h} ,
\end{align}
for each $Y$, $Y' \in \mathcal{h}$, and $Z \in \mathcal{l}$.
See \cite{Radenkovi__2019} for more information about these maps.
To obtain the 2-form Yang-Mills equations, one defines a map $\alpha^*:\mathcal{g} \longrightarrow \mathcal{h}$ in \cite{2002hep.th....6130B} by the rule:
\begin{align}\label{Qw3}
\langle Y, \alpha^*(X)\rangle_\mathcal{h}=\langle \tilde{\alpha}(Y), X\rangle_\mathcal{g}, \ \ \ \ \ \ \ \forall X \in \mathcal{g}, Y \in \mathcal{h}.
\end{align}
Here, to obtain the 3-form Yang-Mills equations, we have to  define a map $\beta^*:\mathcal{h} \longrightarrow \mathcal{l}$ by the rule:
\begin{align}\label{Qw4}
\langle Z, \beta^*(Y)\rangle_\mathcal{l}=\langle \tilde{\beta}(Z), Y\rangle_\mathcal{h}, \ \ \ \ \ \ \ \forall Y \in \mathcal{h}, Z \in \mathcal{l}.
\end{align}
The maps $\sigma$, $\kappa$, $\eta_1$, $\eta_2$, $ \alpha^*$ and $\beta^*$ are very closely linked with our approach to the construction of the 3-form Yang-Mills.

In order to calculate more efficiently, we introduce the component notation.
Given a Lie algebra $\mathcal{g}$, there is a vector space $\Lambda^k (M,\mathcal{g})$
of $\mathcal{g}$-valued differential $k$-forms on the manifold $M$. 
For $A=\sum\limits_{a}A^a X_a \in \Lambda^{k_1} (M, \mathcal{g}) $, $A'=\sum\limits_{b}A'^b X_b \in \Lambda^{k_2} (M, \mathcal{g}) $ for some scalar differential  $k_1$-form $A^a$, and $k_2$-form $A'^b$,and elements $X_a, X_b \in \mathcal{g}$, 
define
\begin{equation*}
A\wedge A' :=\sum\limits_{a,b}A^a \wedge A'^b X_a X_b\ , \quad A \wedge^{\left[, \right]} A' :=\sum\limits_{a,b}A^a \wedge A'^b \left[X_a, X_b\right], \quad dA= \sum\limits_{a} dA^a X_a,
\end{equation*}
then there is an identity
$$A\wedge^{\left[, \right]} A'=A\wedge A'-(-1)^{k_1 k_2}A'\wedge A.$$
In the above, we can also choose $\mathcal{g}$ to be $\mathcal{h}$ or $\mathcal{l}$.
For $B=\sum\limits_{b} B^b Y_b \in \Lambda^{k_1}(M,\mathcal{h})$, $B'=\sum\limits_{b} B'^b Y_b \in \Lambda^{k_2}(M,\mathcal{h})$, $ \forall$ $Y_a$, $Y_b \in \mathcal{h}$, define
$$A\wedge^{\widetilde{\vartriangleright}} B :=\sum\limits_{a,b} A^a \wedge B^b X_a \widetilde{\vartriangleright} Y_b\ ,\quad B\wedge^{\left\{, \right\}} B':=\sum\limits_{a,b} B^a \wedge B'^b \left\{Y_a,Y_b\right\},\ \  \tilde{\alpha} (B) := \sum\limits_{a}B^a \tilde{\alpha}(Y_a).$$  
For $C=\sum\limits_{a}C^a Z_a \in \Lambda^{k_1} (M, \mathcal{l})$, define
$$ A\wedge^{\widetilde{\vartriangleright}} C :=\sum\limits_{a,b} A^a \wedge C^b X_a \widetilde{\vartriangleright} Z_b\ , \ \ B\wedge^{\widetilde{\vartriangleright}'}C:=\sum\limits_{a,b}B^a \wedge C^b Y_a\widetilde{\vartriangleright}' Z_b\ , \ \  \tilde{\beta}(C):=\sum\limits_{a}C^a \tilde{\beta}(Z_a),$$
where $Y_a \widetilde{\vartriangleright}' Z_b = -\left\{\tilde{\beta}(Z_b), Y_a\right\}$ by using \eqref{YZ}.
	
	Furthermore, we have non-degenerate $G$-invariant forms in $\Lambda^k(M, \mathcal{g})$, $\Lambda^k(M, \mathcal{h})$ and $\Lambda^k(M, \mathcal{l})$ induced by $\langle, \rangle_\mathcal{g}$, $\langle, \rangle_\mathcal{h}$ and $\langle, \rangle_\mathcal{l}$ and we denote them by $\langle,\rangle$. Then we have
	$$\langle A, A'\rangle := \sum\limits_{a,b}A^a \wedge A'^b \langle X_a, X_b\rangle_\mathcal{g}, \quad \langle B, B'\rangle := \sum\limits_{a,b}B^a \wedge B'^b \langle Y_a, Y_b\rangle_\mathcal{h}, \quad \langle C, C'\rangle := \sum\limits_{a,b}C^a \wedge C'^b \langle Z_a, Z_b\rangle_\mathcal{l},$$
		where $C'=\sum\limits_{b}C'^b Z_b \in \Lambda^{k_2} (M, \mathcal{l})$,  and $A$, $A'$, $B$, $B'$ are as above.
	There are identities
	$$ \langle A, A'\rangle=(-1)^{k_1 k_2} \langle A', A\rangle, \qquad \langle B, B'\rangle=(-1)^{k_1 k_2} \langle B', B\rangle, \qquad \langle C, C'\rangle=(-1)^{k_1 k_2} \langle C', C\rangle,$$ 
	using the symmetry of $\langle, \rangle_\mathcal{g}$, $\langle, \rangle_\mathcal{h}$ and $\langle, \rangle_\mathcal{l}$.
	
	There is an important proposition in \cite{doi:10.1063/1.4870640}, and we use it to calculate the 3-form Yang-Mills equations.
	\begin{proposition}
		\begin{enumerate}
			\setlength{\itemsep}{6pt}
			\item For $A\in \Lambda^k (M,\mathcal{g})$, $A'\in  \Lambda^{k'}(M,\mathcal{g})$ and $C\in \Lambda^* (M,\mathcal{l})$,
			\begin{equation}\label{AC}
			\tilde{\beta}(A\wedge^{\widetilde{\vartriangleright}} C)=A\wedge^{\widetilde{\vartriangleright}} \tilde{\beta}(C),
			\end{equation}
			\begin{equation}
			A\wedge^{\widetilde{\vartriangleright}} A' = A\wedge A'+(-1)^{kk'+1} A'\wedge A.
			\end{equation}
			
			\setlength{\itemsep}{6pt}
			\item For $A\in \Lambda^k (M, \mathcal{g})$, $B_1 \in \Lambda^{t_1}(M, \mathcal{h})$, $B_2 \in \Lambda^{t_2} (M,\mathcal{h})$ and $W \in \Lambda^* (M, \mathcal{w})$ $(\mathcal{w}=\mathcal{g}, \mathcal{h}, \mathcal{l})$,
			\begin{equation}\label{AW}
			d(A \wedge^{\widetilde{\vartriangleright}} W)=dA \wedge^{\widetilde{\vartriangleright}} W + (-1)^k A \wedge^{\widetilde{\vartriangleright}} dW ,
			\end{equation}
			\begin{equation}\label{BB}
			d(B_1\wedge^{\left\{,\right\}}B_2)=dB_1\wedge^{\left\{,\right\}}B_2+(-1)^{t_1} B_1 \wedge^{\left\{,\right\}}dB_2,
			\end{equation}
			\begin{equation}\label{ABB}
			A\wedge^{\widetilde{\vartriangleright}} (B_1\wedge^{\left\{,\right\}}B_2)=(A\wedge^{\widetilde{\vartriangleright}} B_1)\wedge^{\left\{,\right\}}B_2 + (-1)^{kt_1}B_1\wedge^{\left\{,\right\}}(A\wedge^{\widetilde{\vartriangleright}}B_2),
			\end{equation}
		\end{enumerate}
	 \end{proposition}
	
	 The following propositions provide identities of non-degenerate symmetric $G$-invariant forms in a differential 2-crossed module.
	\begin{proposition}
		For $A_1 \in \Lambda^{k_1}(M, \mathcal{g})$, $A_2 \in \Lambda^{k_2} (M,\mathcal{g})$, $A_3 \in \Lambda^{k_3} (M,\mathcal{g})$, $B_1\in \Lambda^{t_1}(M,\mathcal{h})$, $B_2\in \Lambda^{t_2} (M,\mathcal{h})$, $B_3\in \Lambda^{t_3} (M,\mathcal{h})$, $C_1\in \Lambda^{q_1}(M,\mathcal{l})$, $C_2\in \Lambda^{q_2} (M,\mathcal{l})$, and $C_3\in \Lambda^{q_3} (M,\mathcal{l})$ we have 
		\begin{align}\label{A_1A_2A_3}
		\langle A_1 \wedge^{[ , ]}A_2, A_3\rangle = (-1)^{k_1k_2+1}\langle A_2, A_1 \wedge^{[,]}A_3 \rangle,
		\end{align}
		\begin{align}
		\langle B_1 \wedge^{[,]}B_2, B_3\rangle = (-1)^{t_1t_2+1}\langle B_2, B_1 \wedge^{[,]}B_3 \rangle,
		\end{align}
		\begin{align}
		\langle C_1 \wedge^{[,]}C_2, C_3\rangle = (-1)^{q_1q_2+1}\langle C_2, C_1 \wedge^{[,]}C_3 \rangle.
		\end{align}
		Proof:We can get these identities easily by using \eqref{XXX}, \eqref{YYY} and \eqref{ZZZ}.
		\begin{flushright}
			$\square$
		\end{flushright}
	\end{proposition}

	\begin{proposition}
		For $A \in \Lambda^{k}(M,\mathcal{g})$, $B_1 \in \Lambda^{t_1}(M, \mathcal{h})$, $B_2 \in \Lambda^{t_2}(M, \mathcal{h})$, $C_1 \in \Lambda^{q_1}(M, \mathcal{l})$ and $C_2 \in \Lambda^{q_2}(M, \mathcal{l})$,  we have
		\begin{align}\label{B_1AB_2}
		\langle B_1, A\wedge^{\widetilde{\vartriangleright}} B_2\rangle =(-1)^{t_2(k+t_1)+kt_1+1}\langle B_2, A\wedge^{\widetilde{\vartriangleright}} B_1 \rangle =(-1)^{kt_1+1}\langle A\wedge^{\widetilde{\vartriangleright}} B_1, B_2\rangle,
		\end{align}
		\begin{align}\label{C_1AC_2}
		\langle C_1, A\wedge^{\widetilde{\vartriangleright}} C_2\rangle =(-1)^{q_2(k+q_1)+kq_1+1}\langle C_2, A\wedge^{\widetilde{\vartriangleright}} C_1 \rangle =(-1)^{kq_1+1}\langle A\wedge^{\widetilde{\vartriangleright}} C_1,  C_2\rangle.
		\end{align}
		Proof:We can get those identities easily by using \eqref{YXY'} and \eqref{ZXZ'}.
		\begin{flushright}
			$\square$
		\end{flushright}
	\end{proposition}
	
	
	
	We can define a bilinear map $\overline{\sigma}: \Lambda^{t_1}(M, \mathcal{h}) \times \Lambda^{t_2}(M, \mathcal{h}) \longrightarrow \Lambda^{t_1+t_2}(M, \mathcal{g})$ by 
	$$\overline{\sigma}(B_1,B_2):= \sum\limits_{a,b}B_1^a \wedge B_2^b \sigma (Y_a,Y_b),$$
	for $ B_1= \sum\limits_{a} B_1^a Y_a \in \Lambda^{t_1}(M, \mathcal{h})$ , $ B_2=\sum\limits_{b} B_2^b Y_b \in \Lambda^{t_2}(M, \mathcal{h})$.
	Since $\sigma$ is antisymmetric, we have 
	$$\overline{\sigma}(B_1,B_2)=(-1)^{t_1 t_2 +1}\overline{\sigma}(B_2,B_1).$$
	Let $A=\sum\limits_{c}A^c X_c \in \Lambda^k(M, \mathcal{g})$,
	then
	\begin{align*}
	\langle \overline{\sigma}(B_1,B_2), A\rangle &=\sum\limits_{a ,b, c}B_1^a \wedge B_2^b \wedge A^c \langle \sigma(Y_a,Y_b), X_c \rangle_\mathcal{g}=-\sum\limits_{a ,b, c}B_1^a \wedge B_2^b \wedge A^c \langle Y_a,X_c \widetilde{\vartriangleright} Y_b\rangle_\mathcal{h}\\[2mm]
	&=(-1)^{kt_2+1}\sum\limits_{a ,b, c}B_1^a \wedge A^c \wedge B_2^b  \langle Y_a,X_c \widetilde{\vartriangleright} Y_b\rangle_\mathcal{h}\\[2mm]
	&=(-1)^{kt_2+1} \langle B_1, A\wedge^{\widetilde{\vartriangleright}} B_2\rangle,
	\end{align*}
	by using \eqref{Qw1}, i.e. 
	\begin{align}
	\langle \overline{\sigma}(B_1,B_2), A\rangle = (-1)^{kt_2+1} \langle B_1, A\wedge^{\widetilde{\vartriangleright}} B_2\rangle,
	\end{align}	
	and the following identity holds
	\begin{align}\label{AB_1B_2}
	\langle  A,\overline{\sigma}(B_1,B_2)\rangle = (-1)^{t_1t_2+1} \langle  A\wedge^{\widetilde{\vartriangleright}} B_2,B_1\rangle.
	\end{align}

	Similarly, we can define a bilinear map $\overline{\kappa}: \Lambda^{q_1}(M, \mathcal{l}) \times \Lambda^{q_2}(M, \mathcal{l}) \longrightarrow \Lambda^{q_1+q_2}(M, \mathcal{g})$ by 
		$$\overline{\kappa}(C_1,C_2):= \sum\limits_{a,b}C_1^a \wedge C_2^b \kappa (Z_a,Z_b),$$
		for $C_1= \sum\limits_{a} C_1^a Z_a \in \Lambda^{q_1}(M, \mathcal{l})$ and $C_2=\sum\limits_{b} C_2^b Z_b \in \Lambda^{q_2}(M, \mathcal{l})$.
	Since $\kappa$ is antisymmetric, we have 
		$$\overline{\kappa}(C_1,C_2)=(-1)^{q_1 q_2 +1}\overline{\kappa}(C_2,C_1).$$
	Let $$A=\sum\limits_{c}A^c X_c \in \Lambda^k(M, \mathcal{g}),$$ then
		\begin{align*}
		\langle \overline{\kappa}(C_1,C_2), A\rangle &=\sum\limits_{a ,b, c}C_1^a \wedge C_2^b \wedge A^c \langle \kappa(Z_a,Z_b), X_c \rangle_\mathcal{g}=-\sum\limits_{a ,b, c}C_1^a \wedge C_2^b \wedge A^c \langle Z_a,X_c \widetilde{\vartriangleright} Z_b\rangle_\mathcal{l}\\
		&=(-1)^{kq_2+1}\sum\limits_{a ,b, c}C_1^a \wedge A^c \wedge C_2^b  \langle Z_a,X_c \widetilde{\vartriangleright} Z_b\rangle_\mathcal{l}\\
		&=(-1)^{kq_2+1} \langle C_1, A\wedge^{\widetilde{\vartriangleright}}C_2\rangle,
		\end{align*}
		by using \eqref{Qw2}, i.e.
	\begin{align}
	\langle \overline{\kappa}(C_1,C_2), A\rangle = (-1)^{kq_2+1} \langle C_1, A\wedge^{\widetilde{\vartriangleright}} C_2\rangle,
	\end{align} 
 and there is an identity
	\begin{align}\label{AC_1C_2}
	\langle  A,\overline{\kappa}(C_1,C_2)\rangle = (-1)^{q_1q_2+1} \langle  A\wedge^{\widetilde{\vartriangleright}} C_2,C_1\rangle.
	\end{align}
	
Finally, we define bilinear maps $\overline{\eta_i}: \Lambda^q (M,\mathcal{l}) \times \Lambda^t(M, \mathcal{h}) \longrightarrow \Lambda^{q+t}(M,\mathcal{h})$ by 
	$$\overline{\eta_i}(C, B):= \sum\limits_{a,b} C^b \wedge B^a \eta_i(Z_b, Y_a),  \ \ \ \ i = 1,2$$
for $B_i= \sum\limits_{a} B^a_i Y_a \in \Lambda^{t_i}(M, \mathcal{h})$, $C= \sum\limits_{b} C^b Z_b \in \Lambda^q(M, \mathcal{l})$.
There are two identities
 	\begin{align}\label{B_1B_2C}
 	\langle  B_1 \wedge^{\left\{,  \right\}} B_2, C \rangle  = (-1)^{t_1(t_2 + q) +1}\langle B_2, \overline{\eta_1}(C, B_1) \rangle = (-1)^{t_2 q +1}\langle B_1, \overline{\eta_2}(C, B_2) \rangle,
 	\end{align}
by using \eqref{YY'Z}.

	For $A\in \Lambda^k (M, \mathcal{g})$, $B \in \Lambda^t (M, \mathcal{h})$ and $C \in \Lambda^q(M, \mathcal{l})$, we have
	\begin{align}\label{BAB}
	\langle B, \alpha^*(A)\rangle= \langle\tilde{\alpha}(B), A\rangle,
	\end{align}
	\begin{align}\label{ASD}
	\langle C, \beta^*(B)\rangle= \langle \tilde{\beta}(C), B\rangle,
	\end{align}
	being induced by \eqref{Qw3} and \eqref{Qw4}.
	
	 Up to present,  we have established the algebraic structures of 3-form Yang-Mills theory. Next, we derive the 3-Bianchi-Identities on account of the above properties for Lie algebra valued differential forms.
	 The 3-connection, on $d$-dimensional spacetime manifold $M$ ($d \geq 4$), is described by a 1-form $A$ valued in the Lie algebra $\mathcal{g}$, a 2-form $B$ valued in the Lie algebra $\mathcal{h}$, and a 3-form $C$ valued in the Lie algebra $\mathcal{l}$.  We refer the interested readers to \cite{doi:10.1063/1.4870640} for more details. The corresponding 3-curvature in this case is given by
	\begin{equation*}
	\Omega_1:= dA +A\wedge A, \qquad  \Omega_2:= dB +A\wedge^{\widetilde{\vartriangleright}} B, \qquad \Omega_3:= dC +A\wedge^{\widetilde{\vartriangleright}}C+B\wedge^{\left\{,\right\}}B.
	\end{equation*}
	There are a fake 1-curvature and a fake 2-curvature defined as
	\begin{equation*}
	F_1:=dA +A\wedge A - \tilde{\alpha}(B),
	\qquad
	F_2:=dB +A\wedge^{\widetilde{\vartriangleright}} B - \tilde{\beta}(C).
	\end{equation*}
	
We call the 3-connection $(A,B,C)$ fake 1-flat, if the fake 1-curvature vanishes, i.e. $F_1 = 0$. Similarly, we call the 3-connection $(A,B,C)$ fake 2-flat, if the fake 2-curvature vanishes, i.e. $F_2 = 0$. If the 3-curvature 4-form vanishes, i.e. $\Omega_3=0$, the 3-connection $(A,B,C)$ will be called 3-flat. When $(A,B,C)$ is both fake 1-flat and fake 2-flat, we say that the 3-connection $(A,B,C)$ is fake-flat. Otherwise, the 3-connection $(A,B,C)$ is not fake-flat.
\begin{theorem}\label{theorem}
	If $(A,B,C)$ is any 3-connection on $M$, then its fake 3-curvature $(\Omega_3, F_1, F_2)$ satisfies the {\bfseries 3-Bianchi-Identities}:
	\begin{equation}\label{1A}
		dF_1 + A\wedge^{[,]}F_1=-\tilde{\alpha}(F_2),
		\end{equation} 
		\begin{equation}\label{2A}
		dF_2 + A\wedge^{\widetilde{\vartriangleright}} F_2=(F_1+\tilde{\alpha}(B))\wedge^{\widetilde{\vartriangleright}} B-\tilde{\beta}(\Omega_3-B\wedge^{\left\{,\right\}}B),
		\end{equation}	
		\begin{equation}
		d\Omega_3 + A\wedge^{\widetilde{\vartriangleright}} \Omega_3=(F_1+\tilde{\alpha}(B))\wedge^{\widetilde{\vartriangleright}} C+(F_2+\tilde{\beta}(C))\wedge^{\left\{,\right\}}B+B\wedge^{\left\{,\right\}}(F_2+\tilde{\beta}(C)).
		\end{equation}	
		Proof:
		\begin{align*}
		dF_1&=d(dA+A\wedge A-\tilde{\alpha}(B))=dA\wedge A-A\wedge dA -d(\tilde{\alpha}(B))\\
		&=-A\wedge^{[,]}F_1-\tilde{\alpha}(F_2+\tilde{\beta}(C)),
		\end{align*}
	using $ \alpha (A\wedge^{\widetilde{\vartriangleright}} B)=A\wedge^{\left[,\right]}\tilde{\alpha}(B)$, and
		\begin{align*}
		dF_2&=d(dB+A\wedge^{\widetilde{\vartriangleright}} B-\tilde{\beta}(C))=dA\wedge^{\widetilde{\vartriangleright}} B-A\wedge^{\widetilde{\vartriangleright}} dB-d(\tilde{\beta}(C))\\
		&=(F_1+\tilde{\alpha}(B))\wedge^{\widetilde{\vartriangleright}} B-\tilde{\beta}(\Omega_3-B\wedge^{\left\{,\right\}}B)-A\wedge^{\widetilde{\vartriangleright}} F_2,
		\end{align*}
	using \eqref{AC} and ${\widetilde{\vartriangleright}}$ is homomorphism, and
		\begin{align*}
		d\Omega_3&=d(dC+A\wedge^{\widetilde{\vartriangleright}} C+B\wedge^{\left\{,\right\}}B)=dA\wedge^{\widetilde{\vartriangleright}} C -A\wedge^{\widetilde{\vartriangleright}} dC +d(B\wedge^{\left\{,\right\}}B)\\
		&=(F_1+\tilde{\alpha}(B))\wedge^{\widetilde{\vartriangleright}} C -A\wedge^{\widetilde{\vartriangleright}}\Omega_3+(F_2+\tilde{\beta}(C))\wedge^{\left\{,\right\}}B+B\wedge^{\left\{,\right\}}(F_2+\tilde{\beta}(C)),
		\end{align*} 
		using \eqref{AW},\eqref{BB},\eqref{ABB} and ${\widetilde{\vartriangleright}}$ is homomorphism.
		\begin{flushright}
			$\square$
		\end{flushright}
	\end{theorem}
	When the maps $\tilde{\alpha}$ and $\tilde{\beta}$ are trivial, i.e. $\tilde{\alpha}(B)=0$ and $\tilde{\beta}(C)=0$, we consider the ordinary 3-curvature $(\Omega_1, \Omega_2, \Omega_3)$ satisfying the {\bfseries 3-Bianchi-Identities} which can be derived from Theorem \ref{theorem}
	\begin{equation}\label{11A}
	d\Omega_1 + A\wedge^{[,]}\Omega_1=0,
	\end{equation} 
	\begin{equation}\label{22A}
	d\Omega_2 + A\wedge^{\widetilde{\vartriangleright}} \Omega_2=\Omega_1\wedge^{\widetilde{\vartriangleright}} B,
	\end{equation}	
	\begin{equation}
	d\Omega_3 + A\wedge^{\widetilde{\vartriangleright}} \Omega_3=\Omega_1\wedge^{\widetilde{\vartriangleright}} C+\Omega_2\wedge^{\left\{,\right\}}B+B\wedge^{\left\{,\right\}}\Omega_2.
	\end{equation}	
	
	Generalizing the Yang-Mills action and the 2-form Yang-Mills action in \cite{2002hep.th....6130B}, we write down the following 3-form Yang-Mills action as a function of the 3-connection (A, B, C) in 3-form Yang-Mills gauge theory:
	$$ S=  \int_{M} \langle F_1 , \ast F_1 \rangle +\langle F_2 , \ast F_2 \rangle+ \langle \Omega_3 , \ast \Omega_3 \rangle, $$
    by setting the variation of the action to zero:
	\begin{align*}
	\delta S =2 \int_{M} \langle \delta F_1 , \ast F_1\rangle +\langle \delta F_2, \ast F_2\rangle 	+\langle \delta\Omega_3, \ast \Omega_3\rangle=0.\\
	\end{align*}
	The first section is as follows:
	\begin{align*}
	\langle \delta F_1,\ast F_1\rangle &= \langle \delta(dA + A \wedge A - \tilde{\alpha}(B)), \ast F_1 \rangle\\
	&= \langle \delta A, d\ast F_1+ A\wedge^{[,]}\ast F_1 	\rangle-\langle \delta B, \alpha^*(\ast F_1)\rangle,
	\end{align*}
	using $\delta(A\wedge A) = A \wedge^{[,]} \delta A$, \eqref{A_1A_2A_3} and \eqref{BAB}.
	The second section is as follows :
	\begin{align*}
	\langle \delta F_2, \ast F_2 \rangle &= \langle \delta(dB + A\wedge^{\widetilde{\vartriangleright}} B 	- \tilde{\beta}(C)), \ast  F_2 \rangle\\[2mm]
	&=-\langle \delta A, \overline{\sigma}(\ast F_2, B)\rangle - \langle \delta B,d \ast 	F_2 + A \wedge^{\widetilde{\vartriangleright}} \ast F_2 \rangle - \langle \delta C, \beta^*(\ast F_2)\rangle,
	\end{align*}
    using $\delta(A \wedge^{{\widetilde{\vartriangleright}}} B) = \delta A \wedge^{\widetilde{\vartriangleright}} B + A \wedge^{\widetilde{\vartriangleright}} \delta B$, \eqref{AB_1B_2}, \eqref{B_1AB_2} and \eqref{ASD}.
	 The third section is as follows :
	\begin{align*}
	\langle \delta\Omega_3, \ast \Omega_3\rangle&= \langle \delta(dC + A \wedge^{\widetilde{\vartriangleright}} C + B\wedge^{\left\{,\right\}} B), \ast \Omega_3 \rangle\\
	&= (-1)^{d-1}\langle \delta A, \overline{\kappa}(\ast\Omega_3, C)\rangle-\langle\delta B,\overline{\eta_2}(\ast\Omega_3, B ) +\overline{\eta_1}(\ast\Omega_3, B) \rangle \\
	& +\langle \delta C, d\ast \Omega_3+A\wedge^{\widetilde{\vartriangleright}} \ast\Omega_3\rangle,
	\end{align*}
	using \eqref{AC_1C_2},\eqref{C_1AC_2} and \eqref{B_1B_2C}.
	Thus we have
	\begin{align*}
	\delta S = &2 \int_{M} \langle \delta A, d\ast F_1+ A\wedge^{[,]}\ast F_1 + (-1)^{d-1} \overline{\kappa}(\ast\Omega_3, C)-\overline{\sigma}(\ast F_2, B)\rangle \\[2mm]
	&\ \ - \langle \delta B,d \ast F_2 + A\wedge^{\widetilde{\vartriangleright}} \ast F_2+\overline{\eta_2}(\ast\Omega_3, B ) +\overline{\eta_1}(\ast\Omega_3, B)+\alpha^*(\ast F_1)\rangle\\[2mm]
	&\ \ + \langle \delta C, d\ast \Omega_3+A\wedge^{\widetilde{\vartriangleright}} \ast\Omega_3 - \beta^*(\ast F_2)\rangle.
	\end{align*}
	We see that the variation of the action vanishes for $\delta A $, $\delta B$ and $\delta C$ if and only if the following field equations hold:
	\begin{align}
	&d\ast F_1+ A\wedge^{[,]}\ast F_1 =\overline{\sigma}(\ast F_2, B)+ (-1)^{d} \overline{\kappa}(\ast\Omega_3, C),\\[3mm]
	&d \ast F_2 + A\wedge^{\widetilde{\vartriangleright}} \ast F_2=-\overline{\eta_2}(\ast\Omega_3, B ) -\overline{\eta_1}(\ast\Omega_3, B)-\alpha^*(\ast F_1),\\[3mm]
	&d\ast \Omega_3+A\wedge^{\widetilde{\vartriangleright}} \ast\Omega_3 = \beta^*(\ast F_2).
	\end{align}
	And when the 3-connection $(A, B, C)$ is fake-flat,  the field equations become
	\begin{align}
	&d \ast \Omega_1 +A \wedge^{[,]}  \ast \Omega_1 =  \overline{\sigma}(\ast \Omega_2, B)+ (-1)^d \overline{\kappa}(\ast\Omega_3, C),\\[2mm]
	&d\ast \Omega_2 + A\wedge^{\widetilde{\vartriangleright}} \ast \Omega_2=-\overline{\eta_2}(\ast\Omega_3, B ) -\overline{\eta_1}(\ast\Omega_3, B),\\[2mm]
	&d\ast \Omega_3+A\wedge^{\widetilde{\vartriangleright}} \ast\Omega_3=0.
	\end{align}
	
	Though one may wonder about 4-gauge theory, to the best of our knowledge, it has not been defined yet. The notion of a 3-crossed module, which should be the foundation of 4-gauge theory, has been developed in \cite{AKU, KOO}. 
	Ideally, in higher gauge theory, the Yang-Mills theory may be generalized to a kind of ``$n$-form Yang-Mills theory'' in accordance with the chosen $n$-group structure $(n > 3)$ \cite{Roberts2007TheIA}. There is no doubt that one may encounter a lot of difficulties.
	
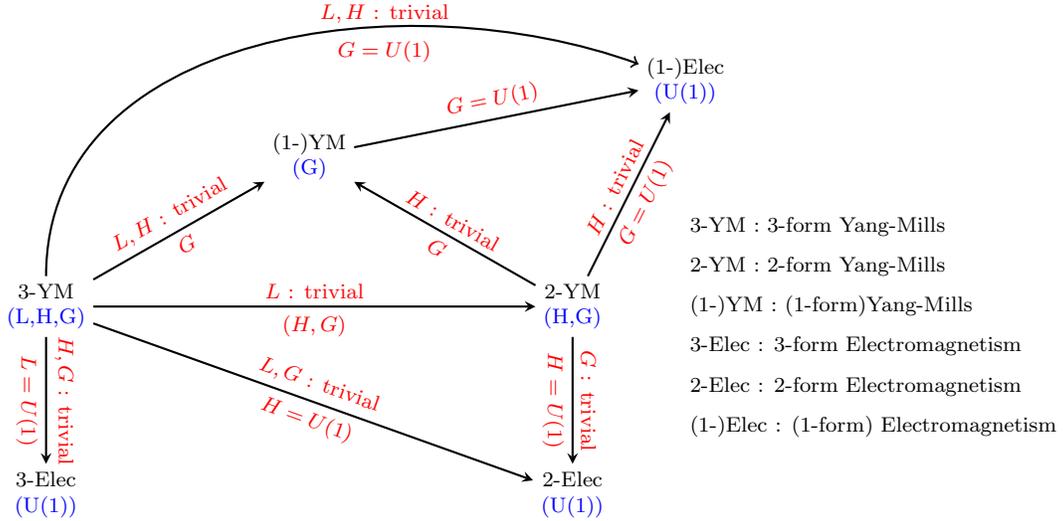
\begin{figure}[H]
	\begin{center}
	\begin{tikzpicture}[node distance=3cm]
	\scriptsize
	\node[align=center] (3YM) at (0,0){\scriptsize 3-YM \\ {\color{blue} (L,H,G)}};
	\node[align=center] (2YM) at (7,0) {\scriptsize 2-YM \\ {\color{blue} (H,G)}};
	\node[align=center] (3ELEC) at (0,-2.5) {\scriptsize 3-Elec \\ {\color{blue} (U(1))}};
	\node[align=center] (2ELEC) at (7,-2.5) {\scriptsize 2-Elec \\ {\color{blue} (U(1))}};
	\node[align=center] (YM) at (3.5,2) {\scriptsize (1-)YM \\ {\color{blue} (G)}};
	\node[align=center] (ELEC) at (8.5,3) {\scriptsize (1-)Elec \\ {\color{blue} (U(1))}};
	\node[color = red] (S) at (4.5,3.9) {\scriptsize $L,H$ : trivial};
	\node[color = red] (S) at (4.5,3.4) {\scriptsize $G = U(1)$};

	\node[align=left] (L1) at (11,-0.5){3-YM : 3-form Yang-Mills \\[2mm] 2-YM : 2-form Yang-Mills \\[2mm] (1-)YM : (1-form)Yang-Mills \\[2mm]  3-Elec : 3-form Electromagnetism \\[2mm]  2-Elec : 2-form Electromagnetism \\[2mm]  (1-)Elec : (1-form) Electromagnetism \\[2mm]};
	
	\draw [arrow] (3YM) -- node[anchor=south,sloped, color=red] {\scriptsize $L$ : trivial}   node[anchor=north, color=red] {\scriptsize $(H,G)$} (2YM);
	\draw [arrow] (3YM) -- node[anchor=south, ,sloped, color=red] {\scriptsize $L, H$ : trivial}   node[anchor=north,  ,sloped, color=red] {\scriptsize $G$} (YM);
	\draw [arrow] (3YM) -- node[anchor=south, ,sloped, color=red] {\scriptsize $H, G$ : trivial}   node[anchor=north, ,sloped, color=red] {\scriptsize $L = U(1)$} (3ELEC);
	\draw [arrow] (3YM) -- node[anchor=south, ,sloped, color=red] {\scriptsize $L, G$ : trivial}   node[anchor=north, ,sloped, color=red] {\scriptsize $H = U(1)$} (2ELEC);
	\draw [arrow] (2YM) -- node[anchor=south, ,sloped, color=red] {\scriptsize $G$ : trivial}   node[anchor=north, ,sloped, color=red] {\scriptsize $H = U(1)$} (2ELEC);
	\draw [arrow] (2YM) -- node[anchor=south, ,sloped, color=red] {\scriptsize $H$ : trivial}   node[anchor=north, ,sloped, color=red] {\scriptsize $G = U(1)$} (ELEC);
	\draw [arrow] (YM) -- node[anchor=south, ,sloped, color=red] {\scriptsize $G = U(1)$}   (ELEC);
	\draw [arrow] (2YM) -- node[anchor=south, ,sloped, color=red] {\scriptsize $H$ : trivial}   node[anchor=north, ,sloped, color=red] {\scriptsize $G$} (YM);
	
	\draw [->,thick] (3YM) to [in = 160, out = 90 ,sloped, color=red] (ELEC);
	
	\end{tikzpicture}
	\end{center}
	\caption{The relationships}\label{figure 7}
\end{figure}

    To summarize, the relationships between the related models are concluded in Figure \ref{figure 7}.
	
\section*{Acknowledgment}
The authors would like to thank referees for useful comments and references.
This work is supported by the National Natural Science Foundation of China (Nos.11871350, NSFC no. 11971322).

	\bibliographystyle{plain}

\begin{thebibliography}{aps}
			
		\bibitem{Baez.2010} J. C. Baez and J. Huerta, An invitation to higher gauge theory, Gen. Relativ. Gravit. 43 (2010) 2335–2392, \href{arXiv:1003.4485 [hep-th]}{arXiv:1003.4485 [hep-th]}.
		
		\bibitem{Baez2005HigherGT} J. C. Baez and U. Schreiber, Higher gauge theory, Commun. Contemp. Math. 431 (2007) 7–30, \href{arXiv:math/0511710}{arXiv:math/0511710}.
		
		\bibitem{Bar}  T. Bartels, Higher gauge theory I : 2-bundles, \href{	arXiv:math/0410328 }{	arXiv:math/0410328 }.
		
		\bibitem{JBUS} J. C. Baez and U. Schreiber, Higher gauge theory: 2-connections on 2-bundles, \href{	arXiv:hep-th/0412325}{	arXiv:hep-th/0412325}.
			
		\bibitem{Faria_Martins_2011} J. F. Martins and R. Picken, The fundamental gray 3-groupoid of a smooth manifold and local 3-dimensional holonomy based on a 2-crossed module, Differ. Geom. Appl. 29 (2011) 179–206, \href{arXiv:0907.2566 [math.CT]}{arXiv:0907.2566 [math.CT]}.
		
		\bibitem{Saemann:2013pca} C. Saemann and M. Wolf, Six-dimensional superconformal field theories from principal 3-bundles over twistor space, Lett. Math. Phys. 104 (2014) 1147–1188, \href{arXiv:1305.4870 [hep-th]}{arXiv:1305.4870 [hep-th]}.
		
		\bibitem{doi:10.1063/1.4870640} W. Wang, On 3-gauge transformations, 3-curvatures, and gray-categories, J. Math. Phys 55 (2014) 043506, \href{arXiv:1311.3796 [math-ph]}{	arXiv:1311.3796 [math-ph]}.
		
		\bibitem{Fiorenza:2012mr}  D. Fiorenza, H. Sati and U. Schreiber, The $E_8$ moduli 3-stack of the C-field in M-theory, Commun. Math. Phys. 333 (2015) 117–151, \href{	arXiv:1202.2455 [hep-th] }{	arXiv:1202.2455 [hep-th] }.
	
		\bibitem{TRMV} T. Radenkovic and M. Vojinovic, Quantum gravity and elementary particles from higher gauge theory, 	Ann. Univ. Craiova Phys. 30 (2020) 74-84, \href{arXiv:2103.08037 [hep-th]}{	arXiv:2103.08037 [hep-th] }.
		
		\bibitem{U} U. Schreiber, From loop space mechanics to nonabelian strings, \href{arXiv:hep-th/0509163}{arXiv:hep-th/0509163}.
		
		
		\bibitem{ACJ} P. Aschieri, L. Cantini and B. Jurco, Nonabelian bundle gerbes, their differential geometry and gauge theory, Commun. Math. Phys. 254 (2005) 367-400, \href{	arXiv:hep-th/0312154}{	arXiv:hep-th/0312154}.
		
		\bibitem{FH} F. Girelli and H. Pfeiffer, Higher gauge theory-differential versus integral formulation, J.Math.Phys. 45 (2004) 3949–3971.
		
		\bibitem{USJ} H. Sati, U. Schreiber and J. Stasheff, $L_{\infty}$-algebras and applications to string- and Chern-Simons $n$-transport, Quantum field theory: competitive models, Springer (2009) 303–424, \href{arXiv:0801.3480 [math.DG]}{arXiv:0801.3480 [math.DG]}.
		
		\bibitem{HSG} H. Sati, Geometric and topological structures related to M-branes, Proc. Symp. Pure Math. 81 (2010) 181-236, \href{arXiv:1001.5020 [math.DG]}{arXiv:1001.5020 [math.DG]}.
		
		\bibitem{SP} S. Palmer, Higher gauge theory and M-theory, \href{	arXiv:1407.0298 [hep-th]}{	arXiv:1407.0298 [hep-th]}.
		
		\bibitem{HP} H. Pfeiffer, Higher gauge theory and a non-Abelian generalization of 2-form electrodynamics, Ann. Phys. 308 (2003) 447-477, \href{arXiv:hep-th/0304074}{arXiv:hep-th/0304074}.
			
    	\bibitem{Kalb:1974yc} M. Kalb and P. Ramond, Classical direct interstring action, Phys. Rev. D. 9 (1974) 2273–2284.
	
    	\bibitem{YZ} Y. Zunger, p-Gerbes and extended objects in string theory, \href{arXiv:hep-th/0002074}{arXiv:hep-th/0002074}.	
	
	    \bibitem{10.1007/BF01889624} M. Henneaux and C. Teitelboim, p-form electrodynamics, Found. Phys. 16 (1986) 593–617.
	
	    \bibitem{EE} E. Esmaeili, p-form gauge fields: charges and memories, \href{arXiv:2010.13922 [hep-th]}{arXiv:2010.13922 [hep-th]}.
	
	   \bibitem{YTMM} Y. Tanizaki and M. Ünsal, Modified instanton sum in QCD and higher-groups, J. High. Energy. Phys. 03 (2020) 123, \href{arXiv:1912.01033 [hep-th]}{arXiv:1912.01033 [hep-th]}.
	
	  \bibitem{Mathai_2006} V. Mathai and D. Roberts, Yang–Mills theory for bundle gerbes, J. Phys. A-Math. Theor. 39 (2006) 6039–6044, \href{arXiv:hep-th/0509037}{arXiv:hep-th/0509037}.
	
   	\bibitem{pittphilsci10008} M. Vákár, Principal bundles and gauge theories, Universiteit Utrecht, \href{http://philsci-archive.pitt.edu/10008/}{http://philsci-archive.pitt.edu/10008/}.
	
		\bibitem{2002hep.th....6130B} J. C. Baez, Higher Yang-Mills theory, \href{arXiv:hep-th/0206130}{arXiv:hep-th/0206130}.
		
		\bibitem{Beaz} J. C. Baez and A. D. Lauda, Higher-dimensional algebra V: 2-groups, Theor. Appl. Categ. 12 (2004)
		423–491, \href{arXiv:math/0307200}{arXiv:math/0307200}.
		
		\bibitem{Crans} J. C. Baez and A. S. Crans, Higher-dimensional algebra VI: lie 2-algebras, Theor. Appl. Categ. 12 (2004) 492–528, \href{arXiv:math/0307263}{arXiv:math/0307263}.
		
		\bibitem{Brown} R. Brown and P. J. Higgins, On the connection between the second relative homotopy groups of some related spaces, P. Lond. Math. Soc. 36 (1978) 193–212.
		
		\bibitem{MORTON2020103548}  J. C. Morton and R. Picken, 2-group actions and moduli spaces of higher gauge theory, J. Geom. Phys. 148 (2020) 103548, \href{arXiv:1904.10865 [math-ph]}{arXiv:1904.10865 [math-ph]}.
		
			\bibitem{Gastel} A. Gastel, Canonical gauges in higher gauge theory, Commun. Math. Phys. 376 (2020) 1053–1071, \href{arXiv:1810.06278 [math-ph]}{arXiv:1810.06278 [math-ph]}.
			
			\bibitem{TRMV1} T. Radenkovic and M. Vojinovic, Gauge symmetry of the 3BF theory for a generic Lie 3-group, \href{arXiv:2101.04049 [hep-th]}{arXiv:2101.04049 [hep-th]}.
			
		\bibitem{YHMNRY} Y. Hidaka, M. Nitta and R. Yokokura, Global 3-group symmetry and 't Hooft anomalies in axion electrodynamics, J. High. Energy. Phys. 01 (2021) 173, \href{arXiv:2009.14368 [hep-th]}{arXiv:2009.14368 [hep-th]}.
		
		\bibitem{TRMV3} T. Radenkovic and M. Vojinovic, Hamiltonian analysis for the scalar electrodynamics as 3BF theory, Symmetry.  Integr. Geom. 12, 620 (2020), \href{arXiv:2004.06901 [gr-qc]}{arXiv:2004.06901 [gr-qc]}.
		
		\bibitem{Kamps20022groupoidEI} K. H. Kamps and T. Porter, 2-groupoid enrichments in homotopy theory and algebra, K-theory. 25 (2002) 373–409.
		
		\bibitem{Mutlu1998FREENESSCF} A. Mutlu, T. Porter and R. Brown, Freeness conditions for 2-crossed modules and complexes, Theor. Appl. Categ. 4 (1998) 174–194.
		
		\bibitem{Porter10thecrossed} T. Porter, The crossed menagerie: an introduction to crossed gadgetry and cohomology in algebra and topology, \href{https://ncatlab.org/timporter/files/menagerie11.pdf}{https://ncatlab.org/timporter/files/menagerie11.pdf}.
		
			
			
		\bibitem{E} G. J. Ellis, Homotopical aspects of Lie algebras, J. Aust. Math. Soc. A. 54 (1993) 393–419.
		
		\bibitem{Martins:2010ry} J. F. Martins and A. Mikovic, Lie crossed modules and gauge-invariant actions for 2-BF theories, Adv. Theor. Math. Phys. 15 (2011) 1059–1084, \href{arXiv:1006.0903 [hep-th]}{arXiv:1006.0903 [hep-th]}.
		
		\bibitem{Radenkovi__2019} T. Radenković, M. Vojinović, Higher gauge theories based on 3-groups, J. High. Energy. Phys. 10 (2019) 222, \href{arXiv:1904.07566 [hep-th]}{arXiv:1904.07566 [hep-th]}.
		
		\bibitem{AKU} Z. Arvasi, T. S. Kuzpinari and E. Ö. Uslu, Three crossed modules, Homology. Homotopy. Appl. 11 (2009) 161–187, \href{	arXiv:0812.4685 [math.CT]}{	arXiv:0812.4685 [math.CT]}.
		
		\bibitem{KOO} T. S. Kuzpınarı, A. Odabaş and E. Ö. Uslu, 3-crossed modules of commutative algebras, \href{	arXiv:1003.0985 [math.CT]}{	arXiv:1003.0985 [math.CT]}.
		
	    \bibitem{Roberts2007TheIA} D. Roberts and U. Schreiber, The inner automorphism 3-group of a strict 2-group, J. Homotopy. Relat. Str. 3 2008 193-245, \href{	arXiv:0708.1741 [math.CT]}{	arXiv:0708.1741 [math.CT]}.
	
	
\end{thebibliography}
	
\end{document}